# Beyond CNNs: Exploiting Further Inherent Symmetries in Medical Image Segmentation

Shuchao Pang, Anan Du, Mehmet A. Orgun, *Senior Member, IEEE*, Yan Wang, *Senior Member, IEEE*, Quan Z. Sheng, *Member, IEEE*, Shoujin Wang, Xiaoshui Huang, and Zhenmei Yu

*Abstract*— Automatic tumor or lesion segmentation is a crucial step in medical image analysis for computer-aided diagnosis. Although the existing methods based on Convolutional Neural Networks (CNNs) have achieved the state-of-the-art performance, many challenges still remain in medical tumor segmentation. This is because, although the human visual system can detect symmetries in 2D images effectively, regular CNNs can only exploit translation invariance, overlooking further inherent symmetries existing in medical images such as rotations and reflections. To solve this problem, we propose a novel group equivariant segmentation framework by encoding those inherent symmetries for learning more precise representations. First, kernel-based equivariant operations are devised on each orientation, which allows it to effectively address the gaps of learning symmetries in existing approaches. Then, to keep segmentation networks globally equivariant, we design distinctive group layers with layer-wise symmetry constraints. Finally, based on our novel framework, extensive experiments conducted on real-world clinical data demonstrate that a Group Equivariant Res-UNet (named GER-UNet) outperforms its regular CNN-based counterpart and the state-of-the-art segmentation methods in the tasks of hepatic tumor segmentation, COVID-19 lung infection segmentation and retinal vessel detection. More importantly, the newly built GER-UNet also shows potential in reducing the sample complexity and the redundancy of filters, upgrading current segmentation CNNs and delineating organs on other medical imaging modalities.

## I. Introduction

MEDICAL image analysis has been playing a growing and critical role in the whole clinical diagnosis process, especially with the rapid development of the medical imaging technology [1-2]. In order to address the new challenges this development has brought on and to support the clinical diagnosis process more effectively, there is an urgent need for the development of automatic and accurate medical image segmentation technologies, exhibiting a lower false positive rate and false negative rate. In particular, tumor (or lesion) segmentation in CT volumes is one of the most challenging tasks in medical image segmentation, because of the following reasons: (1) the existing low contrast between tumors and their surrounding tissues with similar appearances; (2) the unpredictability of tumors in location, shape, size, number from different patients; (3) the intensity dissimilarity within different parts in a tumor; and (4) the availability of only limited medical data and potential manual errors in pixel-level annotations.

Currently, Convolutional Neural Networks (CNNs) have been utilized as a proven and effective approach to address the above challenges [3-4]. Based on CNNs, many new segmentation models have been proposed, e.g., the patch-based model, the image-based model, the non-local- or the attention- or the context aggregation-based models, and the 3D CNN model [5, 38-41]. Leaving aside the pros and cons of those models, we find that they have partly improved the performance of medical tumor segmentation by designing novel network architectures. Most importantly, the common trait among these advanced segmentation models is that their performance and reliability heavily depend on regular CNN operations. Nevertheless, as can be observed, in contrast to the human visual system which can detect symmetries existing in images effectively, basic convolutional operations in regular CNNs can only exploit translational invariance, while overlooking further inherent symmetries in medical images. For example, in the test images in Fig. 1, we can observe that the upper well-trained

Manuscript received xxxxxx; revised xxxxxx; accepted xxxxxx. This work was supported in part by the International Macquarie University Research Excellence Scholarship (iMQRES: 2018150) and by the National Natural Science Foundation of China (No. 61472416). This work also benefited from with the assistance of resources and services from the National Computational Infrastructure (NCI), which is supported by the Australian Government. (Shuchao Pang and Anan Du contributed equally; Corresponding author: Mehmet A. Orgun.)

Shuchao Pang is with the School of Cyber Science and Engineering, Nanjing University of Science and Technology, Nanjing 210094, Jiangsu, China and also with the School of Computing, Macquarie University, Sydney, NSW 2109, Australia (e-mail: pangshuchao@njust.edu.cn).

Anan Du is with the School of Electrical and Data Engineering, University of Technology Sydney, NSW 2007, Australia (e-mail: anan.du@student.uts.edu.au).

Mehmet A. Orgun, Yan Wang, Quan Z. Sheng are with the School of Computing, Macquarie University, Sydney, NSW 2109, Australia (e-mail: {mehmet.orgun, yan.wang, michael.sheng}@mq.edu.au).

Shoujin Wang is with the School of Computing, Macquarie University, Sydney, NSW 2109, Australia and also with the Data Science Institute, University of Technology Sydney, NSW 2007, Australia (e-mail: shoujin.wang@mq.edu.au).

Xiaoshui Huang is with Shanghai AI Laboratory, Shanghai 200433, China (e-mail: huangxiaoshui@pjlab.org.cn).

Zhenmei Yu is with the School of Data and Computer Science, Shandong Women's University, Jinan 250014, China (e-mail: zhenmei_yu@sdwu.edu.cn).

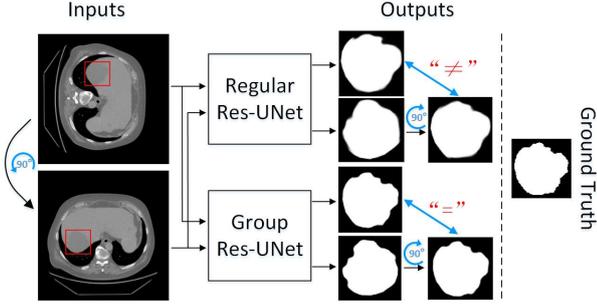

Fig. 1. Equivariance visualization between regular CNNs and group CNNs on a test CT slice and its rotated version. Note that only segmented tumor regions are shown as outputs for clarity.

Res-UNet model based on regular CNNs cannot generate consistent predictions for tumor regions when only rotating the same test slice, which indicates defects of regular CNNs in medical image segmentation tasks.

In order to equip regular CNNs with the capability of exploiting more symmetry properties, three types of representative solutions have been reported. **Firstly**, the widely used data augmentation, as a common and effective method, has been proposed to obtain approximate invariance towards various transformations [6]. Data augmentation can constrain the network training process as a whole to learn more filters for different transformations and thus improve performance, however, (Shortcoming 1) *the size of learned CNN models grows bigger leading to a higher redundancy of kernels and a higher overfitting risk.* More importantly, (Shortcoming 2) *this soft constraint does not guarantee invariance of well-trained CNNs on the test data or even on the training data, as shown in Fig. 1*. **Secondly**, instead of implicitly learning symmetries by rotating input images, existing rotation equivariant networks with explicitly constraining feature maps can maintain multiple rotated feature maps at each layer and are easy to be implemented [7]. Consequently, (Shortcoming 3) *the computation of these rotation operations largely increases memory requirements by rotating and repeating original feature outputs at each layer.* **Finally**, exploiting rotation equivariance by acting on filters has become a promising research direction [8-9]. Although rotating convolutional kernels can achieve local symmetries among different orientations at each convolutional layer, (Shortcoming 4) *these solutions generally limit the depth and the global rotation equivariance of networks due to the dimensionality explosion and the exacerbating noise from orientation pooling operations*.

This paper proposes corresponding solutions to the four shortcomings given above, inspired by group equivariant CNNs for general image classification [10]. For Shortcoming 1, *we build a symmetry group by incorporating translations, rotations and reflections together to significantly increase the utilization of each kernel and reduce the numbers of filters*. For Shortcoming 2, *we design strong layer-wise constraints with rigorous mathematical foundations to guarantee networks be equivariant at each layer*. For Shortcoming 3, *we perform our equivariant transformations on filters, rather than on feature maps, in order to reduce memory requirements*. For Shortcoming 4, *our proposed group equivariant layers and modules can keep networks globally equivariant in an end-to-end fashion. Moreover, they can be stacked into deeper architectures for various vision segmentation tasks, with a negligible computational overhead over regular CNN-based counterparts*.

By integrating the above solutions to the four shortcomings, we propose a novel group equivariant segmentation framework by designing new mathematically sound additional group layers and modules, according to inherent symmetries in medical images. As shown in Fig. 1, the bottom Group Res-UNet based on our proposed framework consistently predicts two liver tumor regions with rotated equivariant results and achieves more accurate segmentation performance.

The main contributions of this work are summarized below:

- We propose novel group equivariant segmentation CNNs going beyond the concept of regular CNNs. Accordingly, more inherent symmetries existing in medical images are captured at each layer in more effective ways.
- The proposed group layers can significantly improve the degree of weight sharing and increase expressive capacity of segmentation networks. More importantly, the group layers of our novel framework can keep the whole group CNNs globally equivariant for segmentation.
- This work reveals a common bottleneck of the current segmentation networks, that is, *localizing tumors is easy, but delineating tumor boundaries is difficult*. To mitigate this problem, all the proposed group layers can be easily generalized in these modern CNN-based segmentation architectures with better performance.

Our empirical study shows that (1) group segmentation CNNs significantly outperform the baseline regular CNN-based counterparts; (2) the proposed GER-UNet achieves the state-of-the-art performance on the real-world clinical hepatic tumor data based on different evaluation metrics; (3) two additional sets of experiments further demonstrate the generalization capability of the proposed method without the corresponding hyper-parameter adjustments for COVID-19 lung infection segmentation on CT data and retinal blood vessel detection on fundus images, compared to scene segmentation methods and specialized retinal vessel detection methods, respectively; and (4) similarly, the proposed method also shows potential in delineating organs on other medical imaging modalities, including prostate segmentation on MRI data and thyroid segmentation on ultrasonography (US) data.

The remainder of this paper is structured as follows: we first review the related work in Section II. Then, we discuss our proposed group equivariant segmentation framework in detail in Section III. Section IV presents experimental settings and performance comparison between our framework and the state-of-the-art methods. In Section V, we further demonstrate the generalization capability of our proposed framework on two more challenging tasks and on two different medical imaging modalities. Finally, in Section VI, we provide a summary of our contributions and also point out several promising future research directions.

## II. RELATED WORK

In this section, we briefly review the equivariance of regular CNNs and introduce a new symmetry group concept. Then, data augmentation is discussed as a widely used approach. Furthermore, we discuss some recent works in rotation equivariant networks and novel group equivariant CNNs.

### A. Equivariant Property & Symmetry Group

We usually regard CNNs having translation invariance because each weight-shared feature kernel can detect the same object that could appear in any position of the whole input image [11]. In other words, when shifting the input image, regular CNNs can give a corresponding shift in the output feature maps, which is also called translational equivariance. Nevertheless, many images, including medical ones, exhibit not only translation equivariance but also rotation and reflection symmetries [12]. The current CNN models lack the ability to exploit such equivariant properties and have to learn more convolutional kernels with more training data to make up for the shortcomings. If we simply use 8 rotated/reflected images as the training input, and train separate networks using different rotated/reflected images, and ensemble them to serve as a general model to address it, it would be an expensive solution and be hard to approximate in an equivariant manner. If we could replace the traditional single translational symmetry by a symmetry group, which covers more equivariant properties such as translations, rotations and reflections, to convolute each input data, and thus this would lead to more powerful and predictable performance in various vision tasks. Therefore, the current regular CNNs can be regarded as a special case of group CNNs.

### B. Data Augmentation & Rotation Equivariant Networks

Data augmentation is a widely used technique to train a more robust neural network model for real applications [6]. In essence, data augmentation methods mainly rely on random transformations on original data to increase the amount of data. Although such augmented data do help improve performance, they require a larger model capacity to save such copies of each feature filter for all the potential transformations. Actually, this soft constraint hardly guarantees equivariant properties on test data, even on training data, as shown in Fig. 1 and Fig. 5, where data augmentation methods including rotations and reflections were also used on the training data. In addition, for the same problem, Dieleman *et al*. [7] proposed a rotation equivariant network by directly rotating each feature map at every layer, but this would substantially increase memory requirements. Recently, some rotated equivariant networks [8-9] rotate convolutional kernels to achieve local symmetries among all the orientations. As a result, the network architectures are usually very shallow and the noise generated from the intermediate layers makes it difficult to maintain global equivariance.

### C. Group Equivariant CNNs

In a groundbreaking work [10], Cohen and Welling introduced group equivariant convolutional neural networks by exploiting symmetry groups for general image classification. In particular, the group convolution operation can increase the representation capacity of the network without increasing the number of parameters. Spurred by this theory, there are a growing numbers of works for various computer vision applications [12-15]. For example, in terms of segmentation tasks, Veeling *et al*. [12] directly utilized the classification network proposed in [10] to distinguish each input patch cropped from the whole image for segmentation. Furthermore, Xiaomeng Li *et al*. [13] proposed an automatic skin lesion segmentation method. Although the method proposes a deeply supervised learning model, it does not give an accurate interpretation with rigorous mathematical proofs. Similarly, Winkens *et al*. [14] proposed to improve semantic segmentation performance by exploiting rotation and reflection symmetries, but there is no detailed method illustration or sufficient experimental verification in their paper. Winkels, and Cohen also utilized the equivariant CNNs for pulmonary nodule detection in CT scans [15].

In contrast to the works discussed above [12-15], our proposed segmentation framework extends the theory of group equivariance [10] from image classification to semantic segmentation, by further devising group equivariant Up-sampling, Output and Skip Connection modules for feature decoding and fusion. Moreover, our framework belongs to image-based segmentation category as it exploits an efficient and end-to-end segmentation network acting on the whole image.

In the next section, we will discuss the proposed group modules in our framework in detail, and then construct a novel Group Res-UNet for hepatic tumor segmentation and other object segmentation tasks in medical image analysis. For the benefit of researchers working in this domain, we plan to make the implementation details available online upon acceptance[1].

## III. THE PROPOSED GROUP EQUIVARIANT SEGMENTATION FRAMEWORK

This section first introduces mathematical convolution formulas for signal and image processing, and then discusses symmetric properties with kernel-based equivariant operations. Then, we propose several core modules by adding layer-wise symmetry constraints into our framework. Finally, we design a Group Equivariant Res-UNet (named GER-UNet) as an example to illustrate how to make these core modules work together for medical image segmentation tasks with the global equivariance.

### A. Mathematical Convolution

In a groundbreaking work [10], Cohen and Welling introduced group equivariant. In mathematics, the convolution operation is a main tool for signal analysis and processing due to signal attenuation over time. For example, assume that an input signal function $f(t)$ and a time response function $g(t)$ are given, then the output signal at time $T$ is calculated as

---
[1] https://github.com/shuchao1212/GER-UNet

follows:
$$[f * g](T) = \int_{-\infty}^{+\infty} f(t)g(T-t)dt. \quad (1)$$

Therefore, we can observe that the convolution operation consists of two parts: a function rollover (from $g(t)$ to $g(-t)$) (including a further sliding ($g(T-t)$)) and an integral (or weighted sum). Based on this theory, the current CNNs also exploit it from the continuous form to the discrete form below:
$$[f * w_i^{(t)}](x) = \sum_{y \in \mathbb{Z}^2} \sum_{k=1}^{K^{(t-1)}} f_k(y) w_{i,k}^{(t)}(x-y), \quad (2)$$

where $f: \mathbb{Z}^2 \to \mathbb{R}^{K^{(t-1)}}$ is the input function at the $t^{th}$ layer which means that the stack of feature maps $f$ outputted at the $(t-1)^{th}$ layer returns a $K^{(t-1)}$ vector at each pixel coordinate $(u, v) \in \mathbb{Z}^2$; similarly, $w_i^{(t)}: \mathbb{Z}^2 \to \mathbb{R}^{K^{(t-1)}}$ is the $i^{th}$ convolutional kernel function at the $t^{th}$ layer. Therefore, for the translation equivariance of regular CNNs, we can see that the translation followed by a convolution is the same as a convolution followed by a translation [10]:
$$\begin{aligned}\left[[L_t f] * w_i^{(t)}\right](x) &= \sum_{y \in \mathbb{Z}^2} \sum_{k=1}^{K^{(t-1)}} f_k(y-t) w_{i,k}^{(t)}(x-y) \\ &= \sum_{y \in \mathbb{Z}^2} \sum_{k=1}^{K^{(t-1)}} f_k(y) w_{i,k}^{(t)}((x-t)-y) \\ &= \left[L_t [f * w_i^{(t)}]\right](x), \quad (3)\end{aligned}$$

where $L_t$ is a translation operator by $y \to y + t$ and $f * w_i^{(t)}$ is also a function on $\mathbb{Z}^2$. However, this property from regular CNNs is not equivariant to a rotation operator $L_r$. On the contrary, this process has to be done by rotating the kernel $w_i^{(t)}$,
$$\left[[L_r f] * w_i^{(t)}\right](x) = \left[L_r \left[f * [(L_r)^{-1} w_i^{(t)}]\right]\right](x), \quad (4)$$

where it is shown that rotating the input feature maps $f$ and then convoluting with a filter kernel $w_i^{(t)}$ is the same as the rotation operation by $L_r$ of the convolution between the original input $f$ and the inverse-rotated filter kernel $(L_r)^{-1} w_i^{(t)}$.

Analogously, the process can also be performed for reflection operators with the above formula. Therefore, to achieve these kinds of goals without additionally learning rotated and reflected copies of the same filter by utilizing more training data, we introduce group operations on symmetry groups to replace these conventional operations in regular CNNs, which can equip CNNs with more equivariant properties.

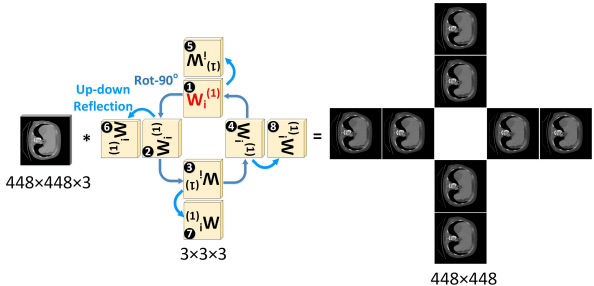

Fig. 2. Visually convolutional representation of the Group Input Layer in our framework.

### B. Core Modules in the Proposed Framework

**The Group Input Layer.** In our framework, all the network operations are based on the same symmetry group, which consists of translations, rotations by multiples of $\pi/2$ and reflections. So, all the convolutional input comes into groups of $\|G = \{g\}\| = 8$ which corresponds 4 pure rotations and their own roto-reflections. Among all the group convolution operations $G$, only the first layer is applied on the original input images, which is called the Group Input Layer. Therefore, in this $\mathbb{Z}^2 \to G$ convolution process, we convolute the input image with 8 rotated and reflected versions of the same kernel, e.g., $w_i^{(1)}$ in Equation (5). The whole group input layer is visually demonstrated in Fig. 2 where the input image is of size $448 \times 448 \times 3$ and the kernel size is $3 \times 3 \times 3$. In addition, the parameters stride=1 and padding=1 are also set to control the size of output like regular CNNs.
$$[f * w_i^{(1)}](g) = \sum_{y \in \mathbb{Z}^2} \sum_{k=1}^{K^{(0)}} f_k(y) w_{i,k}^{(1)}(g^{-1}y), \quad (5)$$

where $K^{(0)}$ is the number of channels from input images. Note that, in this layer, the input image $f: \mathbb{Z}^2 \to \mathbb{R}^{K^{(0)}}$ and the filter $w_i^{(1)}: \mathbb{Z}^2 \to \mathbb{R}^{K^{(t-1)}}$ all belong to functions of $\mathbb{Z}^2$, but $f * w_i^{(1)}$ is a function on group $G$. And, the equivariance (under translations, rotations and reflections) of the group input layer can be derived by analogy to Equation (3), with any predefined group operators, e.g., $L_r: y \to ry$ as follows:
$$\begin{aligned}\left[[L_r f] * w_i^{(1)}\right](g) &= \sum_{y \in \mathbb{Z}^2} \sum_{k=1}^{K^{(0)}} f_k(r^{-1}y) w_{i,k}^{(1)}(g^{-1}y) \\ &= \sum_{y \in \mathbb{Z}^2} \sum_{k=1}^{K^{(0)}} f_k(y) w_{i,k}^{(1)}((r^{-1}g)^{-1}y) \\ &= \left[L_r [f * w_i^{(1)}]\right](g). \quad (6)\end{aligned}$$

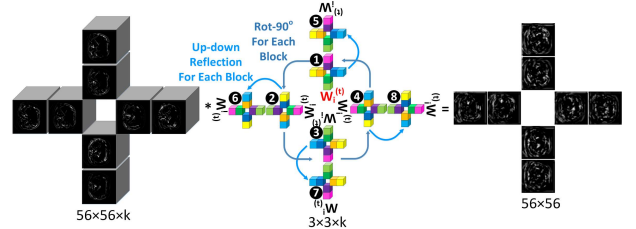

Fig. 3. Visually convolutional representation of the Group Hidden Layer in our framework.

**The Group Hidden Layer.** Different to the group input layer, the next hidden layers are all operated on feature map groups based on the outputs from the previous layer. So, we call this $G \to G$ convolution process as the Group Hidden Layer. Because different kernels are designed for different orientations of input group feature maps, we need to convolute the input feature maps with 8 rotated and reflected symmetric group operations $G$ for each orientation. The details about this group hidden layer can be further understood in the following equation and the visually convolutional representation as shown in Fig. 3. Therefore, all the layers and filters after the first group input layer are all functions on $G$.
$$[f * w_i^{(t)}](g) = \sum_{h \in G} \sum_{k=1}^{K^{(t-1)}} f_k(h) w_{i,k}^{(t)}(g^{-1}h). \quad (7)$$

Similar to Equation (5), the equivariance (under translations, rotations and reflections) of the group hidden layer can be derived with any predefined group operators $L_r: h \to rh$ as follows:
$$\left[[L_r f] * w_i^{(t)}\right](g) = \sum_{h \in G} \sum_{k=1}^{K^{(t-1)}} f_k(r^{-1}h) w_{i,k}^{(t)}(g^{-1}h)$$

$$= \sum_{h \in G} \sum_{k=1}^{K^{(t-1)}} f_k(h) w_{i,k}^{(t)}((r^{-1}g)^{-1}h)$$
$$= \left[ L_r [f * w_i^{(t)}] \right](g). \quad (8)$$

**The Group Up-sample Layer.** Like traditional upsampling operations in regular CNNs, e.g., FCNs [16], we can also interpolate pixels into input feature maps with different modes such as the nearest and bilinear ones. In order to increase the size of outputs after each upsampling operation, we replace the traditional upsampling process only on an orientation by a Group Up-sample Layer over all the 8 orientations. To this end, we could upsample the symmetric group feature maps from each orientation respectively, or concatenate all the group feature maps at the current layer and interpolate them all at once, and then separate them into 8 orientations again. Therefore, the upsampling operation is equivariant by acting on each group feature position from group equivariant feature maps. Note that this process is performed in sequence for keeping the group up-sample layer equivariant as well.

**Group Skip Connections.** Skip connections have been proven effective in segmentation networks for recovering detailed features by combining encoder and decoder feature maps, for example, UNet [17] and FCNs [16]. Because of regular skip connections acting on two individual feature output blocks, we extend this operation on group outputs over all the orientations. In order to connect two sets of group convolutional outputs from the encoder and the decoder stages, we add or concatenate them together, following each orientation. In this way, our Group Skip Connections can obtain more details from each symmetry property, leading to more accurate segmentation predictions. Meanwhile, the sum of two group equivariant feature maps is also group equivariant.

**The Group Output Layer.** This layer is the last layer in our framework to generate the final segmentation score results, called the Group Output Layer. Furthermore, this is a $G \to \mathbb{Z}^2$ aggregation process, which aims to aggregate all the group segmentation outputs over various orientations. More importantly, this aggregation layer is quite critical to keep our group segmentation framework equivariant over different rotations and reflections. To this end, we adopt globally average pooling over each orientation to obtain the equivariance for segmentation tasks. In our medical tumor segmentation tasks, we utilize the group output layer to transform all the orientation channels $f(h)$ into a single pixel-wise 2D predicted output map $f(x)$.

$$f(x) = \frac{1}{\|G\|} \sum_{h \in G} f(h). \quad (9)$$

Note that we ignore the group max-pooling layer in our segmentation framework due to its significant reduction in resolution of feature maps, which is not conducive for segmentation [18], instead we use our group hidden layer with different strides. In addition, other batch normalization operations and non-linear pointwise activations (e.g., ReLU) are also locally equivariant on each symmetry group $G$, which can allow all these group layers to be stacked for much deeper group CNN segmentation models with global equivariance.

*C. A GER-UNet Model*

To evaluate the performance of our proposed group equivariant segmentation framework, we design a GER-UNet model for medical tumor segmentation, which is based on ResNet blocks and our proposed core modules. The whole architecture of our proposed GER-UNet is shown in Fig. 4, where all the convolutions, batch normalizations, activation operations and other layers are constructed by our group equivariant counterparts. The architecture consists of one group input layer, eight ResNet blocks, numerous group hidden layers, three group up-sample layers, one group output layer and others for the pixel-wise segmentation task.

IV. EXPERIMENTS AND ANALYSIS

Extensive experiments are conducted on a real clinical hepatic tumor CT dataset, to answer the following questions: 1) is the whole segmentation framework equivariant to rotation? 2) can the group equivariant CNN model outperform its regular CNN counterpart for segmentation? and 3) is such a simple GER-UNet model better or more competitive than the state-of-the-art methods in this task?

*A. Experimental Settings*

**Datasets & Evaluation Metrics.** To evaluate our proposed group equivariant segmentation framework, a public and available liver tumor segmentation dataset [19] is used in all the experiments. This dataset includes 131 contrast-enhanced abdominal 3D CT scans, which are collected from 131 studied subjects who were suffering from various hepatic tumor diseases and different medical imaging acquisition devices from around the world. In addition, of all the CT slices, there are only 7,190 slices with tumor information. To comprehensively compare tumor segmentation results among different segmentation methods, we use the most complete evaluation criteria in image segmentation tasks [20-21], including Dice, Hausdorff distance, Jaccard, Precision (called positive predictive value), Specificity (called true negative rate) and F1 score. It should be kept in mind that, the smaller the Hausdorff distance, the better the segmentation results; and for the other criteria the opposite is the case.

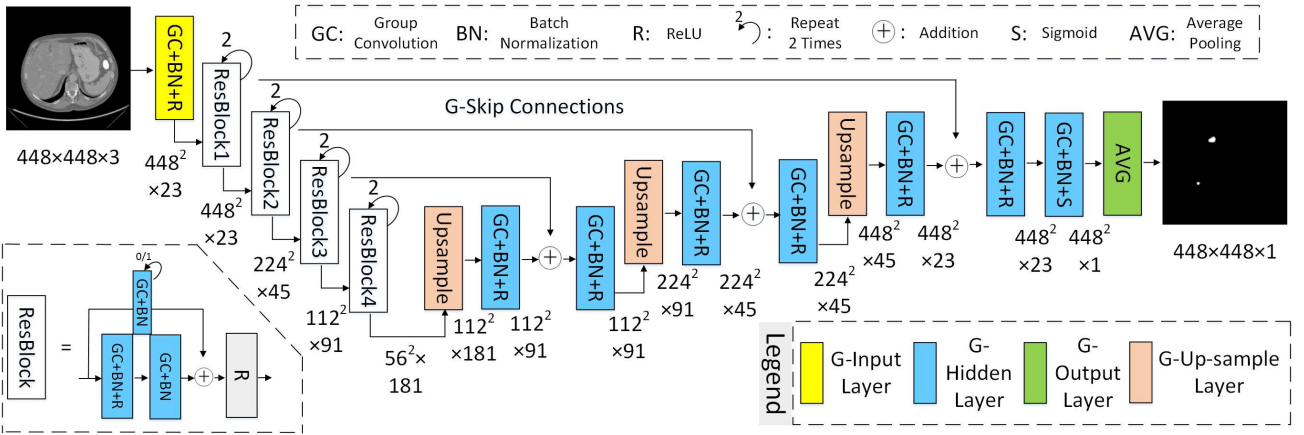

Fig. 4. The architecture of our proposed Group Equivariant Res-UNet (called GER-UNet) for medical image segmentation.

**Parameter Setting.** The dataset is randomly split into 4:1 for model training and testing. All the experiments for all the segmentation methods were conducted on Nvidia Tesla Volta V100 to which access was provided by the National Computational Infrastructure (NCI)[2]. For our GER-UNet and the baseline regular R-UNet models, the batch size is set to 4 for training and the initial learning rate is 2e-4. The learning rate will gradually decrease as the number of training times increases. The training process is performed over 300 epochs, with an early stopping strategy for obtaining the optimal parameters. The basic cross entropy loss function is used to compute the loss errors after each epoch. Also, we adopt the common Adam optimizer to update the whole network parameters. In addition, following the same data preprocessing from [20], all images are normalized into [-1.6, 1.6] before inputted to the training. The widely used data augmentation techniques are also used to train the regular R-UNet model and others, including image contrast transformation in HSV color space, image shift, scale, aspect, rotation, vertical flip and horizontal flip [20]. In other words, the training images contain rotated and reflected images for model training. The other methods chosen for comparison are implemented following the details in the corresponding papers and codes on the same datasets and settings with ours.

**Comparison Methods.** We select 9 state-of-the-art segmentation methods to compare with ours grouped under three categories: (1) U-Net and its variants: U-Net [17], Attention UNet [21] and Nested UNet [22]; (2) Context Based Methods: R2U-Net [23], CE-Net [20] and Self-attention model [24]; (3) Attention Based Methods: SENet [25], DANet [18] and CS-Net [5].

### B. Results and Findings on the Hepatic Tumor Dataset

The experimental results are presented in Table I and all discussions and findings will be reported as follows according to the three questions raised in the first paragraph of Section IV.

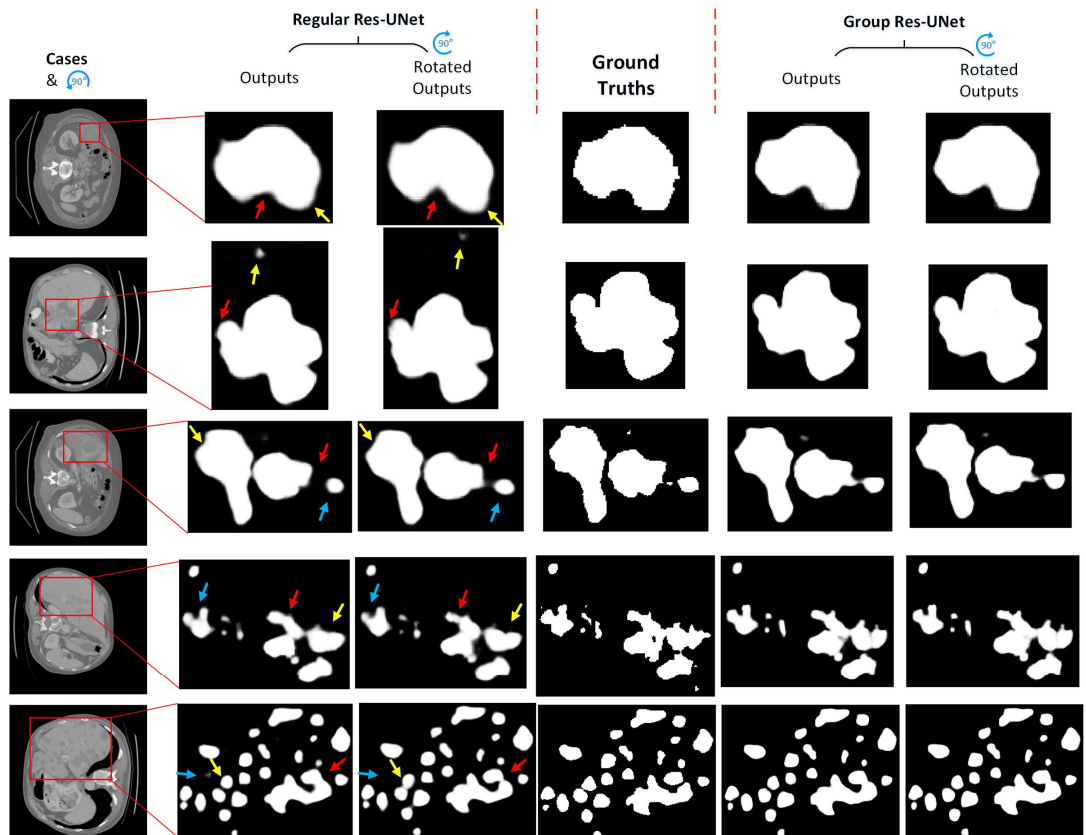

Fig. 5. Equivariant comparison on predicted outputs between regular segmentation CNNs and our group segmentation CNNs on several testing cases. Note that the inputs include the original testing images and their corresponding rotated versions like Fig. 1. In addition, we only show predicted tumor regions due to page limitations and arrows indicate significant differences on both the outputs from regular CNNs.

---

[2] https://opus.nci.org.au/

TABLE I
PERFORMANCE OF SEGMENTATION METHODS ON THE HEPATIC TUMOR SEGMENTATION DATASET, MEASURED BY WIDELY USED EVALUATION METRICS.
NUMBERS IN BOLD REPRESENT THE BEST RESULTS AND UNDERLINED ONES MEAN THE SECOND BEST. NOTE THAT THE HAUSDORFF DISTANCE USES PIXEL
UNITS AND OTHERS %.

| | | Methods/Metrics | Dice | Hausdorff Distance | Jaccard | Precision | Specificity | F1 |
|---|---|---|---|---|---|---|---|---|
| The State-f-the-art | UNet-s | U-Net [17] | 80.27 | 40.13 | 71.30 | 74.95 | 99.86 | 81.68 |
| | | Attention UNet [21] | 81.81 | 34.23 | 73.39 | 77.18 | 99.87 | 83.03 |
| | | Nested UNet [22] | 80.14 | 36.65 | 71.29 | 74.79 | 99.86 | 81.46 |
| | Context | R2U-Net [23] | 78.75 | 35.43 | 69.27 | 72.84 | 99.83 | 80.34 |
| | | CE-Net [20] | 84.43 | 27.13 | 76.75 | 81.34 | 99.91 | 85.35 |
| | | Self-attention [24] | 83.52 | 31.56 | 75.83 | 81.16 | 99.91 | 84.61 |
| | Attention | SENet [25] | 83.42 | 30.45 | 75.58 | 79.95 | 99.90 | 84.43 |
| | | DANet [18] | 85.48 | 27.50 | 78.55 | 84.05 | 99.93 | 86.35 |
| | | CS-Net [5] | 84.12 | 26.42 | 76.21 | 80.23 | 99.90 | 85.11 |
| Ablation Study | | Regular R-UNet | 82.60 | 32.44 | 74.77 | 79.82 | 99.90 | 83.95 |
| | | GER-UNet (ours-w.-*add*) | **86.63** | **24.79** | **80.31** | **87.23** | **99.95** | **87.51** |
| | | GER-UNet (ours-w.-*concat*) | 86.16 | 26.83 | 79.77 | 86.27 | 99.94 | 87.21 |

**Finding 1: GER-UNet has a robust equivariance**

To evaluate the stability of predictions and the equivariance under rotation of the same input, we present a visual analysis shown in Fig. 1 and Fig. 5. Although data augmentation techniques are used on each training image and different at each epoch, the well-trained regular R-UNet gives very different predictions between an original test image and its rotated version, especially, for the boundary regions, which shows defects of current regular CNNs in medical image segmentation. In contrast, the same R-UNet architecture with our proposed group layers is equivariant to the rotation operation on the test image, allowing us to obtain the same predictions by rotating the corresponding output score map. More importantly, by encoding these translation, rotation and reflection equivariances on each symmetry group $G$, the learned group CNN model accurately discriminates object regions with much clearer boundaries, which look more like the corresponding ground truth in Fig. 5. In short, our proposed segmentation framework keeps the learned group CNN model equivariant for input transformations and substantially improves the segmentation performance over its standard CNN counterpart.

**Finding 2: GER-UNet is overwhelmingly superior compared to its regular CNN counterpart**

As mentioned before, regular CNNs can be regarded as a special case of group CNNs because the former has only a single translation equivariance, whereas ours has more equivariant properties. As an ablation study, in order to fairly evaluate the differences between them, we design a novel and standard UNet architecture based on ResNet, called regular R-UNet (or Res-UNet) and then we replace all those basic operations by our group equivariant layers (see Section III). Meanwhile, in order to keep the model parameters consistent between them, we reduce the filter size of each layer to $1/\sqrt{8}$ due to 8 symmetry operations in our group $G$ for each filter, with total 12.14M (ours-w.-*add*) vs 12.12M parameters (Regular R-UNet). As shown in Table I, we observe that our group equivariant segmentation model (GER-UNet) performs consistently better than its regular CNN version (Regular R-UNet). In particular, in terms of the important Dice similarity coefficient and Precision indices, ours is 4.03 percent and 7.41 percent higher than its corresponding Regular R-UNet. This indicates that our novel group equivariant operations significantly improve the performance of medical tumor segmentation in comparison to regular CNN-based layers.

**Finding 3: GER-UNet is more accurate than the state-of-the-art methods**

The current novel and popular segmentation methods build their network architectures deeper and wider based on UNet or FCNs, while also embedding more advanced techniques into networks, as illustrated in the compared methods. To evaluate these state-of-the-art approaches and ours, we have trained and tested them on the same clinical medical tumor dataset. As shown in Table I, the results indicate that our proposed group equivariant GER-UNet performs consistently better than all the compared methods under different evaluation metrics for hepatic tumor segmentation, with the best Dice value with 86.63%, Jaccard index with 80.31%, Precision with 87.23%, F1 score with 87.51% and the shortest Hausdorff Distance with 24.79 pixels. Overall, the average gain of our GER-UNet over all the compared methods (including the baseline Regular R-UNet) achieves clear improvements, which respectively are 4.18% in Dice, 7.40% in Hausdorff Distance, 6.02% in Jaccard, 8.60% in Precision, 0.06% in Specificity and 3.88% in F1 score. This illustrates that our group equivariant segmentation model accurately captures hepatic tumor positions and gives more refined tumor boundary delineations. In other words, it significantly reduces the rates of false positive and false negative results for early medical tumor detection and segmentation. The superior performance of our GER-UNet stems from the increased parameter sharing by encoding more robust symmetric operations for each convolutional filter. Meanwhile, we also observe that our group equivariant model has a faster convergence rate with fewer iterations (about 80 epochs) compared to those of other models (about 300 epochs) based on standard CNNs. For another ablation study, we have updated skip connections by concatenating both feature map blocks from encoding and decoding stages, and it (ours-w.-*concat*) also achieves the best performance as shown in Table I.

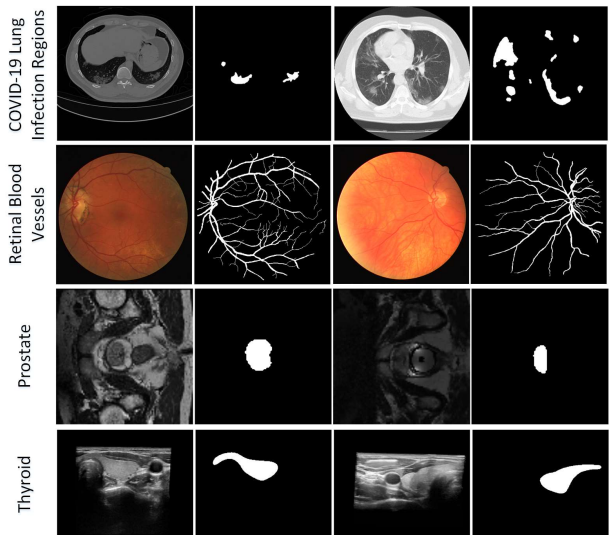

Fig. 6. Several examples we used in our experiments from the lung infection dataset of COVID-19, the retinal vessel detection dataset, the prostate MRI dataset and the thyroid US dataset, respectively.

In short, as discussed above, a simple GER-UNet has already fully demonstrated the potential of our proposed group equivariant segmentation framework. Moreover, we believe that the proposed group network framework and group operations would yield better results when deployed in deeper networks and/or combined with modern popular techniques (e.g., Non-local models, attention schemes and context aggregations).

## V. DISCUSSION

Generalized experiments are implemented to further evaluate the effectiveness and performance of the proposed method on two challenging tasks on real-world medical datasets. One of them is to localize and segment lung infection regions for patients with COVID-19, compared to several popular and state-of-the-art scene segmentation methods. The other one is to detect retinal blood vessels on Fundus images by comparing ours with those specialized retinal vessel segmentation methods. In addition, two different medical imaging modality datasets are also used to further validate the proposed method with the same experimental settings, which are prostate segmentation on MRI data and thyroid segmentation on ultrasonography (US) data, respectively. Furthermore, 95% confidence intervals of our performance results on all the datasets are listed in the supplementary materials and we give comparisons of computation time complexity and number of parameters as well. We should note that there are two basic but important objectives for the test of the generalization capability of a newly proposed method, which are respectively (1) whether the proposed method works well on other medical image segmentation tasks; (2) whether the proposed method works well with the same experimental settings without further tuning them for different tasks and datasets (also called without human interaction); and (3) whether the proposed method shows potential in delineating organs on other medical imaging modalities.

### A. On the Lung Infection Dataset of COVID-19

**Dataset.** This is a new and urgent medical image analysis task under the current influence of COVID-19, which is based on a well-labelled COVID-19 CT dataset as a benchmark to evaluate different annotation methods for lung and infection regions. The dataset includes 20 COVID-19 3D CT scans collected by Ma *et al*. [26]; there are several challenging tasks on this dataset. In this experiment, we choose a challenging and related task with our method to recognize and label infected lung regions automatically, as shown in Fig. 6 with several examples. The most difficult point of this task is that only 4 training cases (20% of all) can be used to train the proposed method and remaining 16 cases (80% of all) are used for testing, which looks like a few-shot or zero-short medical image segmentation task. Therefore, in order to deal with this challenge, we use the open first fold data sets of 5-fold separated combinations to train and test our proposed group equivariant segmentation method (i.e., GER-UNet).

**Results.** As we propose a novel medical image segmentation framework, the implementation details are based on a 2D medical image segmentation architecture. Therefore, an important experiment with the proposed method is a comparison with those popular and new semantic scene segmentation methods despite the fact that 2D medical image segmentation and general scene segmentation belong to two kinds of research problems from two research fields. So, this experiment mainly performs a performance comparison between these robust segmentation methods on the COVID-19 lung infection task to analyze their differences, including the newest benchmark result from 3D nnU-Net.

As can be observed in Table II, we have several revealing findings: **(1) Robust scene segmentation methods cannot exhibit consistent high segmentation performance on COVID-19 lung infection regions.** Moreover, the state-of-the-art method DANet on scene segmentation tasks only achieves 69.19% performance in Dice, whilst the popular and classic DeepLabv3+ and PSPNet give more accurate predictions of lung infections of COVID-19 with respect to Dice of 72.85% and 73.84%. Similar results in this task are also obtained using other evaluation metrics among these scene segmentation methods. **(2) Our group equivariant segmentation method obtains more stable and robust performance under various evaluation performance than those of scene semantic segmentation methods.** For example, the Dice result of ours is 73.93%, the precision of 82.09% and the F1 of 76.27%. Meanwhile, we need to point out that our GER-UNet is trained from scratch while scene segmentation methods are usually pre-trained on ImageNet or have used some pre-trained base networks in their network architectures. And **(3) the 3D medical segmentation method is worse than other 2D ones in this few-shot task.** As we know, 3D medical image segmentation methods are also quite popular in dealing with 3D medical imaging scans because the relationship between the neighboring 2D slices can also be encoded into models for accurate segmentation. However, in the few-shot learning task, the 3D nnU-Net method only achieves 68.08% Dice Coefficient result which is really less than those of 2D segmentation methods. One reason we can think of might be that 3D medical segmentation methods usually include multiple 3D convolutional operations with a larger amount of parameters than 2D convolutional ones, which naturally leads to the overfitting risk when there are only a few training scans.

TABLE II
PERFORMANCE COMPARISON OF SEMANTIC SEGMENTATION METHODS FOR COVID-19 LUNG INFECTION REGIONS BASED ON CT DATA, WHERE WE COMPARE OURS WITH SCENE SEGMENTATION METHODS AND THE OPEN BENCHMARK RESULT OF A 3D METHOD FROM THE AVAILABLE DATASET, MEASURED BY WIDELY USED EVALUATION METRICS. NUMBERS IN BOLD REPRESENT THE BEST RESULTS AND UNDERLINED ONES MEAN THE SECOND BEST. NOTE THAT THE HAUSDORFF DISTANCE USES PIXEL UNITS AND OTHERS %.

| Methods/Metrics | Dice | Hausdorff Distance | Jaccard | Precision | Specificity | F1 |
|---|---|---|---|---|---|---|
| PSPNet [27] | 73.84 | **69.92** | **61.10** | 80.38 | 99.77 | 76.12 |
| DeepLabv3+ [28] | 72.85 | 81.50 | 60.12 | 79.07 | 99.80 | 75.50 |
| DANet [18] | 69.19 | 97.32 | 56.44 | 74.32 | 99.63 | 72.81 |
| 3D nnU-Net [26] | 68.08 | - | - | - | - | - |
| GER-UNet (ours-w.-*add*) | **73.93** | 92.25 | 61.00 | **82.09** | **99.82** | **76.27** |

## B. On the Retinal Vessel Detection Dataset

**Dataset.** The related task is to detect retinal blood vessels from fundus images where the public DRIVE benchmark dataset [29] is used for evaluating different medical image segmentation methods. There are totally 40 2D fundus images in this dataset, a half of which are used for training and the other half for testing. Moreover, in order to annotate retinal blood vessels on each image, two experts have manually labelled them. In our experiments, we also follow the same training and testing sets from the dataset and their labels from the first expert (as the ground truth) to train our GER-UNet and then test the well-trained model on the testing set. As shown in Fig. 6 with several examples, we can observe that the main challenges of this task on DRIVE include low contrast, complex backgrounds, limited training cases and a large proportion of small retinal vessels. All the experimental results are taken from the published papers which proposed the methods that are compared with in our experiments and the evaluation criteria are also the same, including the sensitivity, the accuracy and the AUC (i.e., the area under receiver operation characteristic curve).

**Results.** As a challenging medical image segmentation task, there are many specialized machine learning based methods to automatically detect and segment retinal blood vessels. This experiment is to show the retinal vessel detection performance of our proposed GER-UNet and its generalization capability following the same hyper-parameters from the above experiments without any human intervention. Among these specialized retinal vessel detection and segmentation methods, we have chosen some conventional machine learning methods [30-32] and some newly-proposed deep neural network methods [17, 20, 33-37] for comprehensive comparisons on the DRIVE benchmark dataset.

As shown in Table III, we can observe that (1) overall, the performance gap between the compared methods is very small in terms of the accuracy measure (from 94.34% to 95.80%) and the AUC measure (from 96.01% to 98.02%, except 86.20% of [32]), while the sensitivity gap is relatively large from 72.50% to 85.59% among all these retinal vessel segmentation methods; (2) under the three metrics, deep learning based methods (e.g., 78.18±3.87% in the sensitivity on average) slightly outperform traditional machine learning methods in this task (e.g., 74.42±1.66% in the sensitivity on average); (3) our proposed GER-UNet has a significant improvement in the sensitivity (85.59%) compared to the others and our accuracy (95.65%) and the AUC (97.49%) results are also competitive. In other words, our proposed group equivariant operations can capture more of the small retinal blood vessels with a lower missed detection rate than other specialized methods on fundus images, which also provides further evidence of the importance of exploiting inherent symmetries of medical images in improving the performance of detection and segmentation of regions of interest.

## C. On Other Medical Imaging Modalities

**Datasets.** *The Prostate MRI dataset* is to segment the prostate in transversal T2-weighted MR images, which involves different patients with benign disease and prostate cancer. In addition, all 50 cases come from multiple centers and multiple MRI device vendors, where we randomly select 5 cases as the test case and others for training. *The Thyroid US dataset* is to segment the thyroid in ultrasonography (US). This dataset includes 16 cases with 3D US volumes and all scans contain healthy thyroids, where 13 cases are randomly used for training and 3 cases for testing. Several examples are shown in Fig. 6 and more details about both datasets can be found in [42-43].

TABLE III
PERFORMANCE COMPARISON OF SPECIALIZED RETINAL VESSEL DETECTION AND SEGMENTATION METHODS ON THE DRIVE DATASET WITH WIDELY USED EVALUATION METRICS (%), INCLUDING TRADITIONAL MACHINE LEARNING METHODS AND NEW DEEP NEURAL NETWORK BASED METHODS. NUMBERS in bold REPRESENT THE BEST RESULTS AND underlined ONES MEAN THE SECOND BEST.

| Methods/Metrics (%) | Azzopardi [30] | Roychowdhury [31] | Zhao [32] | HED [33] | U-Net [17] | DeepVessel [34] | Backbone of [20] | CE-Net [20] | Deformable-ConvNet [35] | MSDNet [36] | DUNet [37] | Ours |
|---|---|---|---|---|---|---|---|---|---|---|---|---|
| Sensitivity | 76.55 | 72.50 | 74.20 | 73.64 | 75.37 | 76.03 | 77.81 | <u>81.23</u> | 73.11 | <u>81.23</u> | 79.63 | **85.59** |
| Accuracy | 94.42 | 95.20 | 95.40 | 94.34 | 95.31 | 95.23 | 94.77 | **95.80** | 94.86 | <u>95.78</u> | 95.66 | 95.65 |
| AUC | 96.14 | 96.72 | 86.20 | 97.23 | 96.01 | 97.52 | 97.05 | <u>97.93</u> | 96.93 | 97.87 | **98.02** | 97.49 |

TABLE IV
PERFORMANCE COMPARISON FOR PROSTATE SEGMENTATION ON MRI DATA AND FOR THYROID SEGMENTATION ON ULTRASONOGRAPHY DATA, MEASURED BY COMMONLY USED EVALUATION METRICS, INCLUDING THEIR MEAN VALUES AND 95% CONFIDENCE INTERVALS. NUMBERS IN BOLD REPRESENT THE BEST RESULTS. NOTE THAT THE HAUSDORFF DISTANCE USES PIXEL UNITS AND OTHERS %.

| | Methods/Metrics | | Dice | Hausdorff Distance | Jaccard | Precision | Specificity | F1 |
|---|---|---|---|---|---|---|---|---|
| Prostate (MRI) | U-Net [17] | Mean | 79.72 | **36.00** | 70.13 | 75.91 | 99.40 | 82.41 |
| | | 95% CI | [79.22, 80.23] | [34.93, 37.08] | [69.56, 70.70] | [75.36, 76.46] | [99.39, 99.41] | -- |
| | Regular R-UNet | Mean | 79.38 | 43.18 | 69.49 | 74.58 | 99.28 | 81.73 |
| | | 95% CI | [78.88, 79.88] | [42.05, 44.30] | [68.94, 70.05] | [74.01, 75.15] | [99.27, 99.30] | -- |
| | GER-UNet (ours-w.-*add*) | Mean | **82.35** | 37.93 | **72.75** | **80.04** | **99.47** | **84.39** |
| | | 95% CI | [81.95, 82.76] | [36.74, 39.12] | [72.25, 73.24] | [79.51, 80.56] | [99.46, 99.49] | -- |
| Thyroid (US) | U-Net [17] | Mean | 74.42 | 95.19 | 65.07 | 81.01 | 99.64 | 75.49 |
| | | 95% CI | [74.35, 74.49] | [94.74, 95.64] | [65.01, 65.14] | [80.94, 81.08] | [99.64, 99.65] | -- |
| | Regular R-UNet | Mean | 72.36 | 87.42 | 62.96 | **82.67** | **99.71** | 74.50 |
| | | 95% CI | [72.29, 72.44] | [87.01, 87.82] | [62.89, 63.03] | [82.60, 82.74] | [99.71, 99.71] | -- |
| | GER-UNet (ours-w.-*add*) | Mean | **75.72** | **80.84** | **65.86** | 80.75 | 99.52 | **77.31** |
| | | 95% CI | [75.66, 75.79] | [80.55, 81.13] | [65.80, 65.92] | [80.68, 80.81] | [99.52, 99.53] | -- |
| | Traditional Methods | Dice | Level Set [43] | 71.17 | Graph Cut [43] | 72.77 | Classifier [43] | 64.33 |

**Results.** Note that we regard both experiments as additional generalization capability tests for our algorithm, using the fixed hyper-parameters and models used in the previous experiments. As shown in Table IV, the quantitative results reveal that the proposed method has good generalization capability on other medical image modalities for segmentation. 95% confidence intervals are also provided to further illustrate the robustness of the proposed method in segmenting the prostate on MRI data and the thyroid on US data. Furthermore, compared to traditional methods [43] in segmenting the thyroid on the same US dataset, ours outperforms them as well. In addition, the qualitative comparison is displayed in supplementary materials which also include more details about used cases in experiments.

## VI. CONCLUSION

To learn more precise representations for medical tumor segmentation, we have proposed a novel segmentation framework going beyond regular CNNs by leveraging more inherent symmetries in medical images. To this end, we have developed kernel-based equivariant operations on every orientation, which can also guarantee whole segmentation networks being globally equivariant by encoding layer-wise symmetry constraints at each group layer. This work not only dramatically reduces the redundancy of network filters, but also reveals a common bottleneck of current segmentation networks. Empirical evaluations on real-world clinical data have also shown the superiority of our novel group CNNs for medical tumor segmentation over other recently proposed methods. More importantly, four additional real-world medical segmentation tasks have further demonstrated the effectiveness and generalization capability of the proposed GER-UNet, including COVID-19 lung infection segmentation on CT data, retinal blood vessel detection on fundus images, prostate segmentation on MRI data and thyroid segmentation on ultrasonography data. In addition, our comprehensive experiments with visualization analysis have also revealed several important findings for medical image segmentation.

In the future, we plan to explore 3D medical equivariant segmentation models and develop more inherent symmetries of medical images, for example, reducing the size of rotation angles. Furthermore, we stipulate that our group segmentation framework can be extended into any popular CNN-based segmentation architectures, with the proposed group up-sample layer, the group output layer and group skip connections. We also plan to explore such extended architectures on other challenging medical imaging tasks.

REFERENCES

[1] Isensee, Fabian, Paul F. Jaeger, Simon AA Kohl, Jens Petersen, and Klaus H. Maier-Hein. nnU-Net: a self-configuring method for deep learning-based biomedical image segmentation. *Nature Methods*, vol. 18, no. 2, pp. 203-211, 2021.
[2] Yijie Huang, Qi Dou, Zixian Wang, Lizhi Liu, Ying Jin, Chaofeng Li, Lisheng Wang, Hao Chen, and Ruihua Xu. 3-d roi-aware u-net for accurate and efficient colorectal tumor segmentation. *IEEE transactions on cybernetics*, vol. 51, no. 11, pp. 5397–5408, 2020.
[3] Jianpeng Zhang, Yutong Xie, Pingping Zhang, Hao Chen, Yong Xia, and Chunhua Shen. Light-weight hybrid convolutional network for liver tumor segmentation. In *Proceedings of the 28th International Joint Conference on Artificial Intelligence*, pp. 4271–4277, 2019.
[4] Shuchao Pang, Anan Du, Mehmet A. Orgun, Zhenmei Yu, Yunyun Wang, Yan Wang, and Guanfeng Liu. CTumorGAN: A unified framework for automatic computed tomography tumor segmentation. *European Journal of Nuclear Medicine and Molecular Imaging*, vol. 47, no. 10, pp. 1–21, 2020.
[5] Lei Mou, Yitian Zhao, Li Chen, Jun Cheng, Zaiwang Gu, Huaying Hao, Hong Qi, Yalin Zheng et al. CS-Net: Channel and spatial attention network for curvilinear structure segmentation. In *Proceedings of the International Conference on Medical Image Computing and Computer-Assisted Intervention*, pp. 721–730, 2019.
[6] Dan C. Cireşan, Alessandro Giusti, Luca M. Gambardella, and Jürgen Schmidhuber. Mitosis detection in breast cancer histology images with deep neural networks. In *Proceedings of the International Conference on Medical Image Computing and Computer-assisted Intervention*, pp. 411–418, 2013.
[7] Sander Dieleman, Jeffrey De Fauw, and Koray Kavukcuoglu. Exploiting cyclic symmetry in convolutional neural networks. In *Proceedings of the 33rd International Conference on International Conference on Machine Learning*, pp. 1889–1898, 2016.

[8] Robert Gens, and Pedro M. Domingos. Deep symmetry networks. In *Proceedings of the Advances in Neural Information Processing Systems*, pp. 2537–2545, 2014.

[9] Diego Marcos, Michele Volpi, Nikos Komodakis, and Devis Tuia. Rotation equivariant vector field networks. In *Proceedings of the IEEE International Conference on Computer Vision*, pp. 5048–5057, 2017.

[10] Taco Cohen, and Max Welling. Group equivariant convolutional networks. In *Proceedings of the International Conference on Machine Learning*, pp. 2990–2999, 2016.

[11] Yann LeCun, Yoshua Bengio, and Geoffrey Hinton. Deep learning. *Nature*, vol. 521, no. 7553, pp. 436–444, 2015.

[12] Bastiaan S. Veeling, Jasper Linmans, Jim Winkens, Taco Cohen, and Max Welling. Rotation equivariant cnns for digital pathology. In *Proceedings of the International Conference on Medical Image Computing and Computer-Assisted Intervention*, pp. 210–218, 2018.

[13] Xiaomeng Li, Lequan Yu, Chi-Wing Fu, and Pheng-Ann Heng. Deeply supervised rotation equivariant network for lesion segmentation in dermoscopy images. In *Proceedings of the OR 2.0 Context-Aware Operating Theaters, Computer Assisted Robotic Endoscopy, Clinical Image-Based Procedures, and Skin Image Analysis*, pp. 235–243, 2018.

[14] Jim Winkens, Jasper Linmans, Bastiaan S. Veeling, Taco S. Cohen, and Max Welling. Improved semantic segmentation for histopathology using rotation equivariant convolutional networks, In *Proceedings of the International Conference on Medical Imaging with Deep Learning*, 2018.

[15] Marysia Winkels, and Taco S. Cohen. Pulmonary nodule detection in CT scans with equivariant CNNs. *Medical Image Analysis*, vol. 55, pp. 15–26, 2019.

[16] Jonathan Long, Evan Shelhamer, and Trevor Darrell. Fully convolutional networks for semantic segmentation. In *Proceedings of the IEEE Conference on Computer Vision and Pattern Recognition*, pp. 3431–3440, 2015.

[17] Olaf Ronneberger, Philipp Fischer, and Thomas Brox. U-net: Convolutional networks for biomedical image segmentation. In *Proceedings of the International Conference on Medical Image Computing and Computer-Assisted Intervention*, pp. 234–241, 2015.

[18] Jun Fu, Jing Liu, Haijie Tian, Yong Li, Yongjun Bao, Zhiwei Fang, Hanqing Lu. Dual attention network for scene segmentation. In *Proceedings of the IEEE Conference on Computer Vision and Pattern Recognition*, pp. 3146–3154, 2019.

[19] Patrick Bilic, Patrick Ferdinand Christ, Eugene Vorontsov, Grzegorz Chlebus, Hao Chen, Qi Dou, Chi-Wing Fu et al. The liver tumor segmentation benchmark (lits). *arXiv preprint arXiv:1901.04056*, 2019.

[20] Zaiwang Gu, Jun Cheng, Huazhu Fu, Kang Zhou, Huaying Hao et al. CE-Net: Context encoder network for 2D medical image segmentation. *IEEE Transactions on Medical Imaging*, vol. 38, no. 10, pp. 2281–2292, 2019.

[21] Jo Schlemper, Ozan Oktay, Michiel Schaap, Mattias Heinrich, Bernhard Kainz, Ben Glocker, Daniel Rueckert. Attention gated networks: Learning to leverage salient regions in medical images. *Medical Image Analysis*, vol. 53, pp. 197–207, 2019.

[22] Zongwei Zhou, Md Mahfuzur Rahman Siddiquee, Nima Tajbakhsh, Jianming Liang. Unet++: A nested u-net archi-tecture for medical image segmentation. In *Proceedings of the Deep learning in Medical Image Analysis and Multimodal Learning for Clinical Decision Support*, pp. 3–11, 2018.

[23] Md Zahangir Alom, Chris Yakopcic, Mahmudul Hasan, Tarek M. Taha, and Vijayan K. Asari. Recurrent residual U-Net for medical image segmentation. *Journal of Medical Imaging*, vol. 6, no. 1, pp. 1–16, 2019.

[24] Xiaolong Wang, Ross Girshick, Abhinav Gupta, and Kaiming He. Non-local neural networks. In *Proceedings of the IEEE Conference on Computer Vision and Pattern Recognition*, pp. 7794–7803, 2018.

[25] Jie Hu, Li Shen, Gang Sun. Squeeze-and-excitation networks. In *Proceedings of the IEEE Conference on Computer Vision and Pattern Recognition*, pp. 7132–7141, 2018.

[26] Jun Ma, Yixin Wang, Xingle An, Cheng Ge, Ziqi Yu, Jianan Chen, Qiongjie Zhu et al. Towards efficient COVID-19 CT annotation: A benchmark for lung and infection segmentation. *Medical Physics*, vol. 48, no. 3, pp. 1197–1210, 2021.

[27] Hengshuang Zhao, Jianping Shi, Xiaojuan Qi, Xiaogang Wang, and Jiaya Jia. Pyramid scene parsing network. In *Proceedings of the IEEE Conference on Computer Vision and Pattern Recognition*, pp. 2881–2890, 2017.

[28] Liang-Chieh Chen, Yukun Zhu, George Papandreou, Florian Schroff, and Hartwig Adam. Encoder-decoder with atrous separable convolution for semantic image segmentation. In *Proceedings of the European Conference on Computer Vision*, pp. 801–818, 2018.

[29] Joes Staal, Michael D. Abràmoff, Meindert Niemeijer, Max A. Viergever, and Bram Van Ginneken. Ridge-based vessel segmentation in color images of the retina. *IEEE Transactions on Medical Imaging*, vol. 23, no. 4, pp. 501–509, 2004.

[30] George Azzopardi, Nicola Strisciuglio, Mario Vento, and Nicolai Petkov. Trainable COSFIRE filters for vessel delineation with application to retinal images. *Medical Image Analysis*, vol. 19, no. 1, pp. 46–57, 2015.

[31] Sohini Roychowdhury, Dara D. Koozekanani, and Keshab K. Parhi. Iterative vessel segmentation of fundus images. *IEEE Transactions on Biomedical Engineering*, vol. 62, no. 7, pp. 1738–1749, 2015.

[32] Yitian Zhao, Lavdie Rada, Ke Chen, Simon P. Harding, and Yalin Zheng. Automated vessel segmentation using infinite perimeter active contour model with hybrid region information with application to retinal images. *IEEE Transactions on Medical Imaging*, vol. 34, no. 9, pp. 1797–1807, 2015.

[33] Saining Xie, and Zhuowen Tu. Holistically-nested edge detection. In *Proceedings of the IEEE International Conference on Computer Vision*, pp. 1395–1403, 2015.

[34] Huazhu Fu, Yanwu Xu, Stephen Lin, Damon Wing Kee Wong, and Jiang Liu. Deepvessel: Retinal vessel segmentation via deep learning and conditional random field. In *Proceedings of the International Conference on Medical Image Computing and Computer-assisted Intervention*, pp. 132–139, Springer, Cham, 2016.

[35] Jifeng Dai, Haozhi Qi, Yuwen Xiong, Yi Li, Guodong Zhang, Han Hu, and Yichen Wei. Deformable convolutional networks. In *Proceedings of the IEEE International Conference on Computer Vision*, pp. 764–773, 2017.

[36] Daniël M. Pelt, and James A. Sethian. "A mixed-scale dense convolutional neural network for image analysis." *Proceedings of the National Academy of Sciences*, vol. 115, no. 2, pp. 254–259, 2018.

[37] Qiangguo Jin, Zhaopeng Meng, Tuan D. Pham, Qi Chen, Leyi Wei, and Ran Su. DUNet: A deformable network for retinal vessel segmentation. *Knowledge-Based Systems*, no. 178, pp. 149–162, 2019.

[38] Xi Wang, Hao Chen, Caixia Gan, Huangjing Lin, Qi Dou, Efstratios Tsougenis, Qitao Huang, Muyan Cai, and Pheng-Ann Heng. Weakly supervised deep learning for whole slide lung cancer image analysis. *IEEE Transactions on Cybernetics*, vol. 50, no. 9, pp. 3950–3962, 2019.

[39] Pradeep Kumar Das, Sukadev Meher, Rutuparna Panda, and Ajith Abraham. An efficient blood-cell segmentation for the detection of hematological disorders. *IEEE Transactions on Cybernetics*, pp. 1–12, 2021.

[40] Jie Xue, Kelei He, Dong Nie, Ehsan Adeli, Zhenshan Shi, Seong-Whan Lee, Yuanjie Zheng, Xiyu Liu, Dengwang Li, and Dinggang Shen. Cascaded multitask 3-d fully convolutional networks for pancreas segmentation. *IEEE Transactions on Cybernetics*, vol. 51, no. 4, pp. 2153–2165, 2019.

[41] Xiaofeng Qi, Junjie Hu, Lei Zhang, Sen Bai, and Zhang Yi. Automated segmentation of the clinical target volume in the planning CT for breast cancer using deep neural networks. *IEEE Transactions on Cybernetics*, pp. 1–11, 2020.

[42] Litjens, Geert, Robert Toth, Wendy van de Ven, Caroline Hoeks, Sjoerd Kerkstra, Bram van Ginneken, Graham Vincent et al. Evaluation of prostate segmentation algorithms for MRI: the PROMISE12 challenge. *Medical Image Analysis*, vol. 18, no. 2, pp. 359–373, 2014.

[43] Wunderling, Tom, B. Golla, Prabal Poudel, Christoph Arens, Michael Friebe, and Christian Hansen. Comparison of thyroid segmentation techniques for 3D ultrasound. *In Medical Imaging 2017: Image Processing*, vol. 10133, pp. 1–7, 2017.

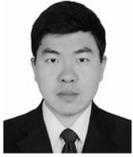

**Shuchao Pang** received his PhD degree in computer science from the Macquarie University, Sydney, Australia in 2020. He is currently a Professor at Nanjing University of Science and Technology, Nanjing 210094, China and also an Honorary Research Fellow with Macquarie University. Before that, he was a Research Associate at the University of New South Wales, Sydney, Australia and a Postdoctoral Researcher at Ingham Institute for Applied Medical Research, Sydney, Australia. His research interests include medical data analysis, deep learning, computer vision and image processing.

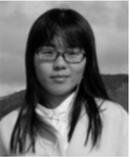

**Anan Du** received her MSc degree in computer application technology from the College of Computer Science and Technology, Jilin University, Changchun, China in 2015. She was an Engineer at China Mobile (HangZhou) Information Technology Co., Ltd., Hangzhou, China from 2015 to 2017. She is currently a PhD student at the School of Electrical and Data Engineering, University of Technology Sydney, NSW 2007, Australia. Her current research interests include 3D point cloud processing, computer vision and deep learning.

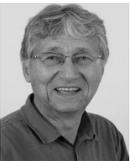

**Mehmet A. Orgun** received the BSc and MSc degrees in computer science and engineering from Hacettepe University, Ankara, Turkey, in 1982 and 1985, respectively, and the PhD degree from the University of Victoria, Canada, in 1991. He is currently a Professor with Macquarie University, Sydney, Australia. His research interests include knowledge discovery, multiagent systems, trusted systems, image processing and temporal reasoning. He is a senior member of the IEEE.

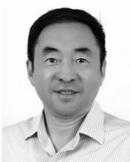

**Yan Wang** received the BEng, MEng, and DEng degrees in computer science and technology from the Harbin Institute of Technology (HIT), P. R. China, in 1988, 1991, and 1996, respectively. He is currently a Professor with the School of Computing, Macquarie University, Sydney, Australia. His research interests include recommender systems, social computing, trust computing and service computing. He is a senior member of the IEEE.

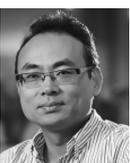

**Quan Z. Sheng** received the PhD degree in computer science from the University of New South Wales, Sydney, Australia, in 2006. He is a Professor with the School of Computing, Macquarie University. His research interests include Internet of Things, big data analytics, distributed computing, and pervasive computing. He is the recipient of the ARC Future Fellowship in 2014, the Chris Wallace Award for Outstanding Research Contribution in 2012, and the Microsoft Research Fellowship in 2003. He is the author of more than 350 publications. He is a member of the ACM and IEEE.

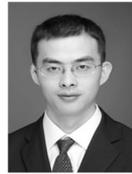

**Shoujin Wang** received his PhD Degree in Data Science from the University of Technology Sydney in 2019. He is currently a Lecturer at the Data Science Institute, University of Technology Sydney, Sydney, NSW 2007, Australia. His research interests include data mining, machine learning and recommender systems. He has published more than 20 research papers in these areas at prestigious journals and conferences. He has served as a program committee at several premier conferences including AAAI, IJCAI, and a reviewer for prestigious journals including machine learning, IEEE Transactions on Cybernetics.

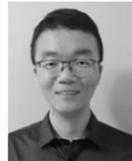

**Xiaoshui Huang** received his PhD from the University of Technology Sydney in 2019. He is currently a Research Scientist at the Shanghai AI Laboratory, Shanghai 200433, China. Before that, he was Postdoctoral Research Associate at the University of Sydney and visiting fellow at the University of Technology Sydney. His research interests are computer vision and machine learning.

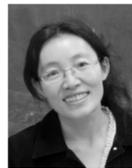

**Zhenmei Yu** received the BSc and MSc degrees in computer science and engineering from Shandong Normal University, Jinan, China, in 1995 and 2008, respectively. She is currently a Professor at the School of Data and Computer Science, Shandong Women's University, Jinan 250014, China. Her research interests include computer education, algorithm design and analysis, data processing and application.